# Electroluminescence in photovoltaic cell

Antonio Petraglia and Valerio Nardone
*Second University of Naples, Department of Environmental Science, Via Vivaldi, 43, I-81100 Caserta, Italy
(e-mail antonio.petraglia@unina2.it)*

Photovoltaic cells are one of the major optoelectronics applications of pn junctions; they transform visible radiation energy into electrical energy. When used the other way around, by driving current through them, they emit radiation. This phenomenon is called electroluminescence (EL), it is the basis of light emitting diodes and semiconductor lasers. The emitted radiation from crystalline silicon photovoltaic cells can not be seen with the naked eye because they emit in the infrared (IR) part of the electromagnetic spectrum, in the wavelength range 950-1300 nm, in an area that is invisible to the human eye, which is sensitive to the 390-750 nm range. The emitted EL intensity is related to the number of minority carriers in the base layer, thus giving informations on intrinsic and extrinsic cell parameters influencing them: minority carrier lifetimes, diffusion length, defects, etc. The IR radiation can be detected by infrared digital cameras. This is one of the most powerful technique to check photovoltaic panels and is getting much interest from the research point of view [1].

Electroluminescence detection system setups, including highly sensitive infrared equipment, power supply and automatic measurement and image acquisition are sold at several tens of thousands of euros and are not suitable for use in a teaching laboratory.

However, it is possible to get EL images in a school lab with a small investment of time and money. Here we propose two of them.

The main issues to face are:
1. the small amount of radiation emitted by the cells;
2. the lack of IR sensitivity of commercial digital cameras; indeed, their CCD sensors are sensitive to the near IR (usually till 1000 nm) but they are bundled with an IR filter to overcome a typical blur given by the IR light;

However, with good planning of the experiment and using proper precautions, it is possible to take interesting pictures of photovoltaic cells emitting EL radiation, suitable for further analysis, by using the residual IR sensitivity of the cameras or removing the IR filter.

Things to consider are:
1. the radiation is very weak; experimenters must be able to eliminate interference from any other source of light: ajar windows and doors but also the lights of the instrumentation;
2. to increase the radiation collected by the camera, it is necessary to act manually on several parameters:
    - focal ratio: by using the maximum aperture (lowest *f*) in order to gather as much light as possible; it should be considered, however, that that will make focusing more difficult;
    - exposure time: by using very long exposures. It requires that the machine is placed firmly on a tripod, it is also a good idea to use a cable release or self timer to avoid vibrations;
    - sensitivity: by using high ISO values. Unfortunately, using high ISO values and long exposure times, we get the typical "noise" and the presence of hot pixels (white spots). Experimenter should check the best parameters that allow a good sensitivity without increasing too much the noise; there are also some good software allowing to remove the hot

pixels and improve the noise (the figures below have been processed with the free software Gimp and the GMIC plug-in). Other techniques, borrowed from the astronomical photography, such as stacking, are possible but not necessary;
- In low light the auto-focus of the camera can cause problems; the manual focus, if available, should be chosen.

The supply of the photovoltaic cell is the heart of the whole process: it is necessary to use a power supply capable of providing the needed power to allow the cell to emit an adequate amount of EL radiation. However, there is a limit; in fact, much of the supplied energy is dissipated in heat and the cell tends to heat up quickly. It is possible to get surface temperatures higher than 100 ° C; however, 70-80 ° C should not be exceeded both for safety reasons and because high temperatures could irreversibly compromise the functioning of the cell. It may be useful to make preliminary experiments by powering the cell and monitoring it with the aid of a surface thermometer. For this reason, it is very important to maintain the power just the time necessary to take the photo and wait long enough for the cell to cool, before taking the next picture.

Operationally, we have to:
- feed the cell, which electrically behaves like a diode, by directly biasing it to obtain a low current trough the cell;
- turn off the light and other sources of spurious radiation in the laboratory;
- take the picture;
- turn off the power immediately to avoid unneeded heating;
- process the photo: removing the hot pixels, possibly using a noise filter and increasing the contrast.

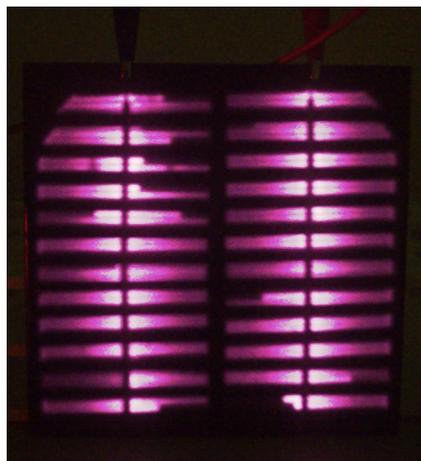

*Figure 1.* Electroluminescence from a didactic photovoltaic cell taken with a commercial digital camera.

For those who have the opportunity to spend a little more time for preparation, there is a second method of capturing images of EL by using a modified webcam. The webcam IR filter is extracted to increase the sensitivity in the area with the highest EL emission; this is a well known technique, in particular in astrophotography. In this case it is possible to shorten the exposure time of at least an order of magnitude, with a consequent improvement in the level of noise, and the ability to shoot video. These improvements are obtained, however, at the expense of a worse resolution.

Figure 1 shows the electroluminescence of a 1,2 W didactic Si cell taken using the first method with a Pentax K100D camera; the latter was chosen because it has a good sensitivity to IR [2]. We set the feed current to 6 A, the feed voltage to 21 V, the exposure to 30 s and sensibility to 3200 ISO.

Emission areas are clearly seen: every junction is divided into strips that are the metal contacts, opaque to light, which collect the current. The bright areas are regions with the highest emission; the basic fact here is that good areas in terms of manufacturing of electronic modules correspond to those of greater electroluminescence.

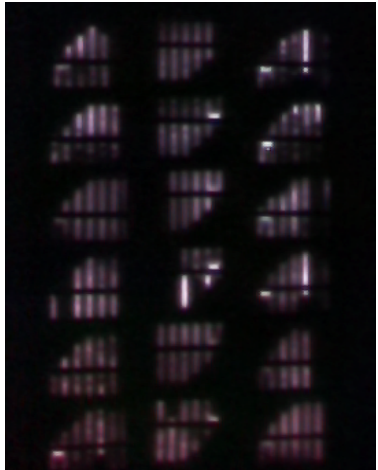

*Figure 2. Picture taken with a modified webcam.*

Figure 2 was taken by using the second method on another Si cell with a different layout geometry. It shows the average of 10 images captured with a webcam logitech-C120 (cost: 10 €). The filter was removed following the instructions taken in [3]. The target was a 0,8 W cell biased at 5 A and 20 V.

The image was taken using the software q*astrocam-g2* (created for astrophotography and very useful for low light situations as these described here). Normal webcam "capture" software can be used with less choices. This second picture has a lower resolution in respect to Figure 1 but a lower noise and a better contrast; however, in both it is possible to discriminate areas of different EL emission and in particular inactive strips in which there is no EL at all.

Further development can include quantitative studies and comparisons between different cell technologies.

Help and discussions with A. D'Onofrio, C. Lubritto and F. Terrasi are gratefully acknowledged.

**References**
[1] Fuyuki T, Kondo H, Yamazaki T, Takahashi Y, and Uraoka Y 2005 Photographic surveying of minority carrier diffusion length in polycrystalline silicon solar cells by electroluminescence, *Appl. Phys. Lett.* **86**, 262108. Bresler M S, Gusev O B, Zakharchenya B P and Yassievich I N 2003 Electroluminescence efficiency of silicon diodes. *Proceedings of the Conference Dedicated to O. V. Losev (1903–1942)*. Nizhni Novgorod, Russia, March 17–20.
[2] http://www.jr-worldwi.de/photo/index.html?ir_comparisons.html
[3] http://www.free-track.net/forum/index.php?showtopic=2219